\documentclass[12pt]{iopart}
\usepackage[margin=3cm]{geometry}  

\expandafter\let\csname equation*\endcsname\relax
\expandafter\let\csname endequation*\endcsname\relax

\usepackage{bm}
\usepackage{graphicx} 
\usepackage{amsmath} 
\usepackage[normalem]{ulem} 
\usepackage{comment}
\usepackage{cite}
\usepackage[dvipsnames]{xcolor}

\newcommand{\Ef}{E^*}

\newcommand{\K}{\mathbf{K}}
\newcommand{\E}{\mathbf{E}}
\newcommand{\B}{\mathbf{B}}
\newcommand{\C}{\mathbf{C}}
\newcommand{\D}{\mathbf{D}}
\newcommand{\upd}{\mathrm{d}}

\begin{document}

\title{Mechanics of elliptical JKR-type adhesive contact}

\author{Andrea Giudici \footnote{Corresponding author: giudici@maths.ox.ac.uk}, Dominic Vella, Ian Griffiths}

\address{Mathematical Institute, University of Oxford, Woodstock Rd, Oxford, OX2 6GG, UK}
\ead{giudici@maths.ox.ac.uk}
\vspace{10pt}

\begin{abstract}

The classic Johnson Kendall Roberts (JKR) theory describes the short-ranged adhesive contact of elastic bodies, but is only valid for axisymmetric contact. A theory for non-axisymmetric contact, which relies on approximating the contact region as an ellipse, was proposed by Johnson and Greenwood (JG). The theory includes the effects of adhesion via Griffith's criterion applied \emph{only} at the semi-major and semi-minor axes of the contact ellipse. Although JG's work is in good agreement with numerical and experimental results for quasi-circular contacts, the agreement worsens as the eccentricity of the contact region increases. In this paper, we show that including the effects of adhesion by instead minimizing the sum of elastic and surface energy yields results that are  in excellent agreement with previous numerical simulations over the full range of contact eccentricities.
\end{abstract}

%
%
%
%
%

\section{Introduction}

The classic Johnson Kendall Roberts (JKR) theory describes the adhesive contact between elastic bodies \cite{johnson1971surface}. It accounts for adhesive interactions via surface energy and is thus only appropriate when the interaction length is small compared to other length-scales in the problem. 
Nevertheless, short-range interactions are common, making the JKR theory extremely useful. For example, JKR is used in the study of adhesion in microelectromechanical systems (MEMS) \cite{mems1,mems2} and biological tissues \cite{bio1,bio2,ciavarella2019role}.

Despite its success, however, the JKR theory only describes axisymmetric contact, and thus cannot describe the general adhesive contact between two bodies --- for example the contact between a cylinder and a sphere. To resolve this limitation, Johnson and Greenwood (JG) \cite{johnson2005approximate} proposed a JKR-type theory (i.e.~for short-range interactions) valid for non-axisymmetric contact. To do so, they noted that, for small-displacements and in the absence of adhesion, the general problem may be reduced to finding the contact region between an ellipsoid and an elastic half-space \cite{barber2000contact}. Since analytical results exist for an elliptical contact region, JG assumed that the contact region remains elliptical even in the presence of adhesion and is entirely specified by the values of the semi-major and semi-minor axes (giving only two degrees of freedom for the solution). To determine these, they then applied Griffith's criterion \cite{fracmec} at the boundary of the semi-major and semi-minor axis of the ellipse  including the effects of adhesion. This allowed them to obtain analytical results that describe the evolution of the indentation, size and shape of the contact ellipse as a function of the applied load. 

Although JG's theory is in excellent agreement with numerics and simulations in the case of weakly elliptical contacts (small eccentricity), numerical studies have shown that the theory underestimates the magnitude of the pull-off force as well as the size of the contact region when the contact ellipse becomes slender (eccentricity close to one) \cite{li2020numerical,jin2011easy}. The relatively large error in the estimation of the pull-off force may be attributed to the incorrect assumption that the contact boundary is elliptical. JG acknowledged that this assumption was likely to be incorrect and it has subsequently been shown both numerically and experimentally that the contact shape is close to being elliptical, but becomes increasingly squared at the edges as the contact becomes more slender \cite{li2020numerical,jin2011easy,sumer2010experimental}. Nonetheless, the discrepancy between the assumed and real contact shapes remains small, suggesting that this difference does not entirely account for the observed error.

Alternative theories and extensions have been proposed to improve upon JG's work. Zini \emph{et al.}~\cite{zini2018extending} derived a theory using the double-Hertz model \cite{greenwood1998alternative}, in which the indenting and adhesive Hertzian pressures act on elliptical domains of different size (but with the same eccentricity), to account for the effects of long-range interactions. However, despite this being a useful model, both the authors of the paper and Greenwood \cite{GreenwoodOn} have expressed worry about the validity of some of the assumptions made. In particular, the authors assume that the eccentricity of the contact region is independent of the load, which appears to be incorrect even for long-ranged interactions. An alternative approach has been proposed by Li and Popov \cite{li2020numerical}, who obtained an approximation for the adhesive elliptical contact by fitting a curve to their numerical simulation data. Their work is useful, but their model lacks a formal physical basis and does not predict the change of eccentricity as a function of load --- something important for applications. 

In this paper, we revisit the problem of elliptical adhesive contact maintaining JG's assumption of an elliptical contact region. However, instead of applying Griffith's criterion at isolated points, we introduce the effect of adhesion by minimizing the sum of the elastic and surface energy of the system. This approach yields results that are in excellent agreement with numerical simulations at various values of eccentricity. Our work indicates that the error in JG's theory arises mostly from the  choice of satisfying Griffith's criterion locally, and not on the assumption of an elliptical contact \emph{per se}. In particular, JG's approach leads them to overestimate the magnitude of the stress singularity at the boundary of contact, thereby underestimating the contact size and the magnitude of the pull-off force.

\begin{figure}[t]
\begin{center}
\includegraphics[width = \textwidth]{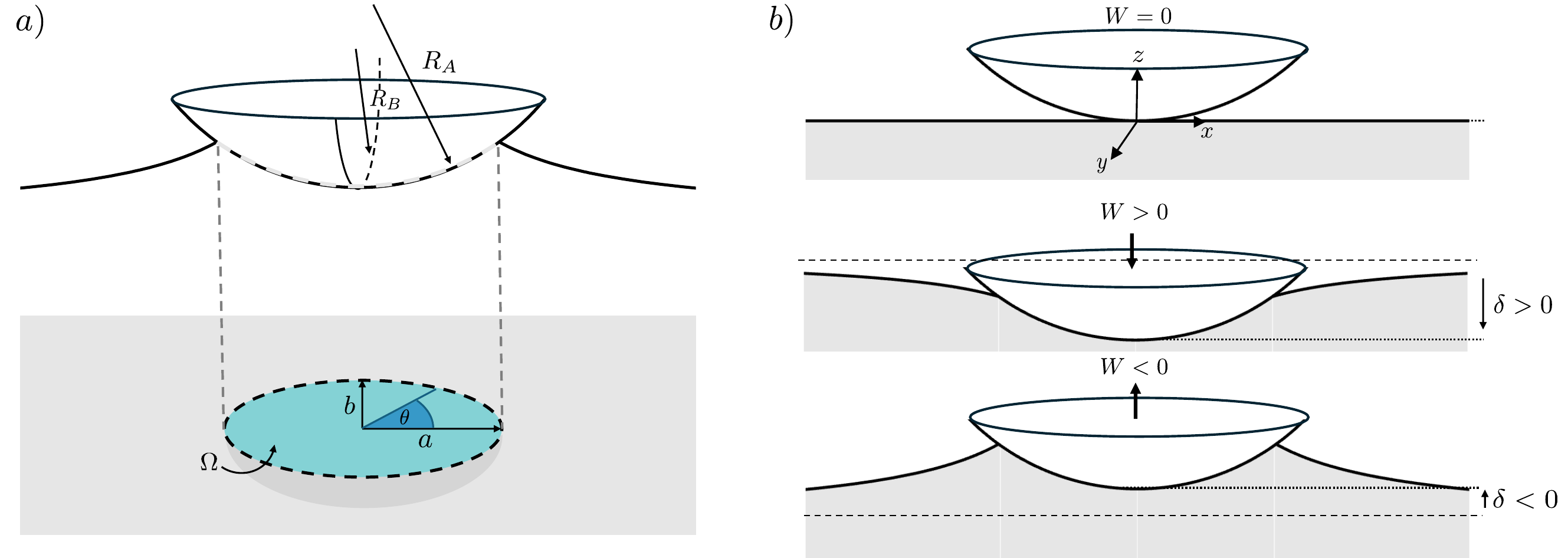}
\caption{a) A sketch of the contact geometry, showing the radii of curvature of the ellipsoid and the contact region. b) Sketch of the hysteretic behaviour of a rigid  ellipsoid as it is loaded and unloaded in contact with an elastic half-space. In the top panel, the ellipsoid comes in contact with the half-space (top) at zero load. It then indents the half-space as it is loaded (middle panel) with depth $\delta$. Finally, even with negative loads the contact is maintained due to adhesive forces, and negative indentations are obtained before the bodies detach (at the pull-off force, $W_{\text{pull off}}$).}
\label{fig:sketch}
\end{center}
\end{figure}

\section{Theory}
We study the adhesive contact between two smooth elastic bodies. When the contact region between the two bodies is small compared to their radii of curvature --- i.e.~the strains and displacements are small --- the problem is equivalent to studying the contact between a stiff (undeformable) ellipsoid coming into adhesive contact with an elastic half-space with given effective stiffness \cite{barber2000contact}. Therefore, without loss of generality, we follow JG's approach and  consider an ellipsoid with radii of curvature $R_A$ and $R_B<R_A$ that indents an elastic half-space whose surface lies on the $x$-$y$ plane. (Coordinates are chosen such that contact first occurs at the point $x=y=z=0$.) The contact region, which we denote $\Omega$, is assumed to be elliptical with semi-major and semi-minor axes $a$ and $b$, respectively. A sketch of the contact geometry is shown in figure \ref{fig:sketch}(a). The load required to push the ellipsoid a depth $\delta$ into the half-space is denoted $W$, while adhesion originates from a surface energy $\Delta \gamma$. 

Solving the adhesive contact problem requires us to find the relationship between the load  $W$, the surface energy $\Delta \gamma$, the contacting geometry $R_A, R_B$, the indentation depth $\delta$, and the size of the contact region (parametrized  by $a$ and $b$). As a first step, we note that, within the contact region, the vertical displacement, $u(x,y)$, at the surface of the elastic half-space caused by the indenting body   can be approximated as parabolic, i.e. 
\begin{eqnarray}
\label{udisp}
    u(x,y)=\delta+\frac{1}{2 R_A} x^2 + \frac{1}{2 R_B} y^2. 
\end{eqnarray}

\noindent From Hertzian contact mechanics \cite{barber2000contact}, it is known that a parabolic indentation is caused by the general  pressure distribution
\begin{eqnarray}
\label{ellipticalpressure}
    p(x,y)=\frac{p_0-\alpha x^2 - \beta y^2}{\sqrt{1-(x/a)^2-(y/b)^2}},
\end{eqnarray}
where $p_0$, $\alpha$ and $\beta$ are all constants that need to be determined. 

Our task is to find $p_0$, $\alpha$, $\beta$, $a$, $b$, and $\delta$ as functions of $W$, $R_A$, $R_B$ and $\Delta \gamma$. This requires six equations, one of which may be obtained by deriving the relationship between indentation and total load $W$ via
\begin{eqnarray}
\label{pint}
    W=\int_{\Omega} p(x,y) \mathrm{~d}x\mathrm{~d}y=2 \pi  c^2 p_0-\frac{2 \pi  c^4 \left(\alpha +\beta  g^2\right)}{3 g}.
\end{eqnarray}
Three further equations come from relating the pressure in eqn \eqref{ellipticalpressure} to the indentation $u$ in eqn \eqref{udisp} and the effective stiffness of the elastic half-space, $E^*$. To write these equations as neatly as possible, and remove any dependence on the stiffness $\Ef$, we non-dimensionalize by making the substitutions:
\begin{eqnarray}
    p_0=\bar{p}_0 \frac{\Ef}{2 b R},\quad \alpha=\bar{\alpha} \frac{\Ef}{2 b R},\quad\beta&=\bar{\beta} \frac{\Ef}{2 b R}. 
\end{eqnarray} 
It is also convenient to characterize the contact geometry using $\lambda=\sqrt{R_B/R_A}$ and $R=\sqrt{R_A R_B}$ while characterizing the contact region using its slenderness $g=b/a$ and the effective contact radius $c=\sqrt{a b}$. Making these substitutions, the relationship between pressure and displacement can be written \cite{johnson2005approximate} as
\begin{gather}
\label{matrixeq}
 \begin{bmatrix} \delta \\ \lambda/(2 R) \\    1/(2 \lambda R) \end{bmatrix}
 =
 \left(\frac{1}{2R} \right) \begin{bmatrix}
   2 \K(e) & c^2/g \textbf{E}(e) & c^2 g \textbf{D}(e) \\
   0       & \textbf{D}(e)+\textbf{C}(e) & -g^2 \textbf{C}(e) \\
   0       & -\textbf{C}(e) &  \textbf{B}(e)+g^2 \C(e) \\
   \end{bmatrix} \cdot \begin{bmatrix} \bar{p}_0 \\ \bar{\alpha} \\ \bar{\beta} \end{bmatrix}.
\end{gather}
where $\K$ and $\E$ are elliptic integrals of the first and second kind respectively, $e=\sqrt{1-g^2}$ is the eccentricity, and we define the functions
\begin{eqnarray}
    \D(e)& \equiv e^{-2}\bigl[\K(e)-\E(e)\bigr],\\
    \B(e)& \equiv \K(e)-\D(e),\\
    \C(e)& \equiv e^{-2}\bigl[\D(e)-\B(e)\bigr].
\end{eqnarray}
 Solving eqn \eqref{matrixeq} for $\delta$, $\bar{\alpha}$ and $\bar{\beta}$, we obtain
\begin{eqnarray}
\label{eq:delta}
    \delta &= \frac{\bar{p}_0 \K(e)}{R}-\frac{c^2 \overline{\alpha } \B(e)}{2 g R}-\frac{c^2 g \overline{\beta } \D(e)}{2
   R},\\
   \label{eq:al}
\bar{\alpha}&=\frac{\lambda ^2 \B(e)+\lambda ^2 g^2
   \C(e)+g^2 \C(e)}{\lambda  \sqrt{g}
   \left[\B(e) \C(e)+\B(e)
   \D(e)+g^2 \C(e) \D(e)\right]},\\
      \label{eq:bet}
\bar{\beta}&=\frac{\lambda ^2
   \C(e)+\C(e)+\D(e)}{\lambda 
   \sqrt{g} \left[\B(e) \C(e)+\B(e)
   \D(e)+g^2 \C(e) \D(e)\right]}.
\end{eqnarray}
In this way, we have now expressed three of the original six unknowns in terms of the remaining three ($c$, $g$ and $\bar{p}_0$).
We can readily find $\bar{p}_0$ by solving equation \eqref{pint}, giving
\begin{eqnarray}
   \bar{p}_0 = \frac{c^2 \left(\overline{\alpha }+g^2 \overline{\beta }\right)}{3 g}+\frac{\sqrt{g} R
   W}{\pi  c \Ef},
   \label{eqn:pValue}
\end{eqnarray}
leaving us only with $c$ and $g$ to be determined. 

\begin{figure}[t]
\begin{center}
\includegraphics[width = \textwidth]{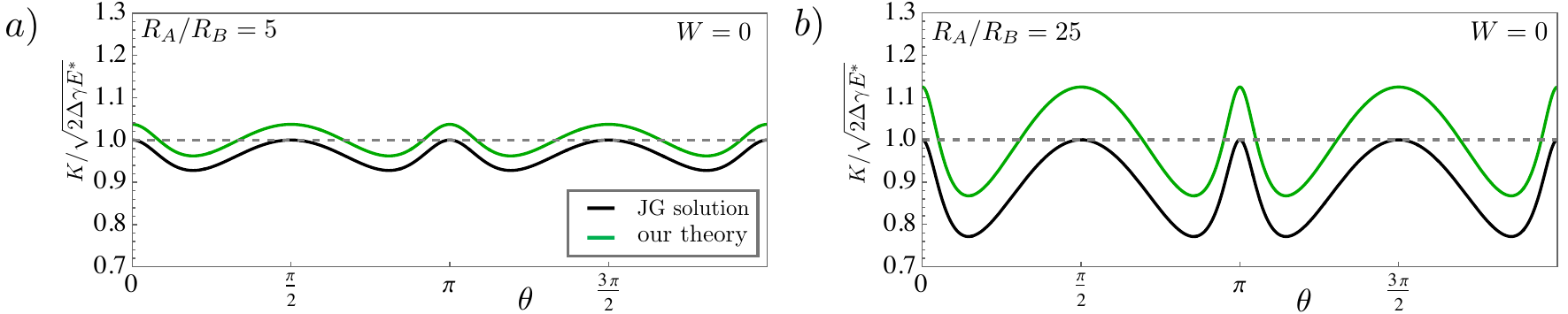}
\caption{ The angular distribution of the normalized stress intensity factor $K/\sqrt{2 \Ef \Delta \gamma}$ around the edge of the contact ellipse obtained by JG and our theory. The dashed line shows the magnitude of the critical stress intensity factor $K_c$ predicted using Griffith's criterion, i.e. the stress intensity factor obtained at the boundary of contact if the contact shape was correct.  Our theory predicts a very similar stress intensity factor distribution as the JG theory, but offset to be slightly larger, and hence on average closer to the dashed line. }
\label{fig:sif}
\end{center}
\end{figure}

A further condition is derived following JG's work and equating the (mode I) stress intensity factor $K$ at the ends of the semi-major and semi-minor axes of contact.  The mode I stress intensity factor describes the strength of the tensile stress singularity at the boundary of a crack or, as in our case, at the boundary of the adhesive contact region \cite{gdoutos2020fracture}.
It is defined as $K= \sqrt{2 \pi} \lim_{r \to 0} \sigma \sqrt{r}$ where $\sigma$ is the stress and $r$ the distance from the boundary. For the correct contact set, $K$ is constant along the boundary. However, the elliptical contact assumption we (and others) have made is only an approximation chosen to facilitate analytical progress; meaning $K$ now varies along the boundary. 
To calculate the stress intensity factor, we identify a point on the boundary via $\theta$, so that $x=a \cos{\theta}$ and $y=b \sin{\theta}$. We then consider being a distance $r$ away from the boundary in the normal direction, with the (unit) normal being $\vec{n}=-(\cos{\theta}/a,\sin{\theta}/b)/\sqrt{(\cos{\theta}/a)^2+(\sin{\theta}/b)^2}$. Using $\sigma=p$ (the pressure is defined in equation \eqref{ellipticalpressure}) and taking the limit as $r \to 0$ we obtain:
\begin{eqnarray}
    K(\theta)=\frac{E^*\sqrt{\pi}}{2 b R}\frac{(\bar{\alpha} a^2 \cos^2 \theta+\bar{\beta} b^2 \sin^2 \theta-\bar{p}_0)}{\left(\cos^2 \theta/a^2+\sin^2 \theta/b^2 \right)^{1/4}},
\end{eqnarray} 
where we remind the reader that $\bar{\alpha}=\alpha(g,\lambda)$, $\bar{\beta}=\beta(g,\lambda)$. JG determined the two unknown parameters $g$ and $c$ by equating the values of $K$ at $\theta=0$ and $\theta=\pi/2$ and, further, by equating this value to that given by Griffith's criterion. Physically, this is equivalent to assuming that the crack propagates first from the semi-major axis of contact. Here, we follow the first of these steps, setting $K(0)=K(\pi/2)$ to give
\begin{eqnarray}
\label{Kmatching}
   (\bar{\alpha} a^2 - \bar{p}_0) \sqrt{ a}=(\bar{\beta} b^2 - \bar{p}_0) \sqrt{b}.
\end{eqnarray}
We can now substitute the expressions for $\bar{\alpha}$ and $\bar{\beta}$, given in eqns \eqref{eq:al} and \eqref{eq:bet}, in \eqref{Kmatching} to obtain the relationship:
\begin{eqnarray}
    W = \frac{\pi   \Ef}{3}c^3\frac{ \left(2 \sqrt{g}+1\right) g^2 \overline{\beta
   }-\left(\sqrt{g}+2\right) \overline{\alpha }}{\left(\sqrt{g}-1\right)
   g^{3/2} R}.
   \label{eqn:gValue}
\end{eqnarray}
Note that we have conveniently expressed $W$ as a function of $g$. Since $W$ (the load) is the relevant control parameter,  $g$ will be used as a parameter to  plot the behaviour of the system.

So far, we have followed the steps from JG's work. To complete their derivation and find the size of the effective contact radius $c$, JG use Griffith's criterion, which predicts delamination occurs when $K$ is larger than the critical value $K_c=\sqrt{2 \Ef \Delta \gamma}$. Since the stress intensity factor is not constant, JG choose to impose Griffith's criterion at the semi-major and semi-minor axis, so that $K(0)=K(\pi/2)=\sqrt{2 \Ef \Delta \gamma}$. However, at these locations the stress intensity factor takes its largest value. Since, for a given elliptical contact, the stress intensity factor increases with the size of the contact radius, choosing the largest possible value of $K$ and equating it to $K_c$ means JG's theory predicts the smallest possible contact area. 

We would like to avoid making a choice on where to estimate the stress intensity factor, and instead satisfy Griffith's criterion in some global sense, over the whole boundary.
In our view, the natural way to do this is to determine the size of the effective contact radius $c$ by minimising the total energy of the system, given by the sum of elastic and surface energy.

Our last step is therefore different from JG's work. We complete our derivation by determining the size of the effective contact radius $c$ using an energetic approach. In particular, we want to write the total energy of the system --- equal to the sum of the elastic energy of the deformation $U_{\text{el}}$ and the energy saved by surface tension via contact $U_{\Delta \gamma}$ --- as a function of the contact radius, i.e. $U_{\text{tot}}(c)=U_{\text{el}}(c)+U_{\Delta \gamma}(c)$. We then find out the size of the contact by minimising the total energy, meaning we need to solve $U'_{\mathrm{tot}}(c)=0$. 
The stress intensity factor (rescaled by the critical value) predicted by this energetic approach is shown by the green curves in figure \ref{fig:sif}. The shapes of the curves are similar to those from JG's theory, but they are shifted upwards. The difference occurs because, for the same elliptical contact size, the global approach sees a smaller stress intensity on the boundary compared with JG's estimate. As a result, a larger contact radius is allowed for the same $\Delta\gamma$, leading to a larger average stress intensity on the boundary.

To evaluate the energy stored in the elastic half-space, we parameterise the pressure and  indentation displacement using a parameter $\xi$.  We let $\tilde{p}(x,y)=\xi p(x,y)$ and $\tilde{u}(x,y)=-\xi u(x,y)$ where $u$ and $p$ are defined in equations \eqref{udisp} and \eqref{ellipticalpressure}, respectively. Note that the minus sign in the latter equation has been added so that positive pressures correspond to displacements \emph{into} the half-space. This parametrisation means that when $\xi=0$ no displacement occurs, while as  $\xi$ increases, so does the deformation of the system until we obtain the final contact deformation at $\xi=1$. By the principle of virtual work, the energy of the deformation is given by
\begin{eqnarray}
    U_{\mathrm{el}}&=\int_{\Omega}\int p(x,y) ~\upd u(x,y) ~\upd A\nonumber\\
    &=\int_{\Omega} \int_0^{1} \tilde{p}(x,y) \frac{\upd \tilde{u}(x,y)}{\upd \xi} ~\upd \xi ~\upd A\nonumber\\
    &=-\int_{\Omega} \int_0^{1} \tilde{p}(x,y) \left(\delta + A x^2 +B y^2 \right)~\upd\xi ~\upd A\nonumber\\
    &= \frac{\pi  c^5 \Ef \left(2 \lambda ^2 \overline{\alpha }+2 g^4 \overline{\beta
   }-g^2 \left(\overline{\alpha }+\lambda ^2 \overline{\beta }\right)\right)}{90
   g^{5/2} \lambda  R^2}-\frac{c^2 W \left(g^2+\lambda ^2\right)}{12 g \lambda 
   R}-\frac{W \delta}{2},
\end{eqnarray}
where in the second step we have used eqn \eqref{udisp}, while in the third step we have evaluated the remaining integral, substituted $A$ and $B$ for $\lambda$ and $R$,  and used equation \eqref{matrixeq}. Note that we have written the dependence of $\delta$ on $c$ arising from equation \eqref{eq:delta} explicitly since this is important later in the minimization. Similarly, we have also had to capture the $W$ dependence explicitly since this is an independent variable.

The final ingredient is the surface energy saved by contact, $U_{\Delta \gamma}=-\Delta \gamma \pi c^2$ where $\pi c^2$ is the contact area of the ellipse. The  total energy, $U_{\mathrm{tot}}=U_{\mathrm{el}}+U_{\Delta \gamma}$ is minimized  with respect to variations in $c$, ($U'_{\mathrm{tot}}(c)=0$), and gives 
\begin{eqnarray}
\label{eqn:cValue}
c\left(\frac{\Ef}{\Delta \gamma  R^2} \right)^{1/3}=  \left(\frac{12 \left(\sqrt{g}-1\right)^2 g^{5/2} \lambda }{f(\lambda,g)-\left(g-\sqrt{g}\right)
   \left(g^{3/2}+\lambda ^2\right) \left(\bar{\beta} g^2-\bar{\alpha}\right)} \right)^{1/3},
\end{eqnarray}
where
\begin{align}
    f(\lambda,g)=   \frac{\lambda
   \left[\left(\sqrt{g}-1\right) \left(\bar{\alpha}
   \B(e)+\bar{\beta} g^2
   \D(e)\right)+\K(e) \left(\bar{\beta}
   g^2-\bar{\alpha} \sqrt{g}\right)\right]}{\left[\bar{\beta}
   \left(2 \sqrt{g}+1\right) g^2-\bar{\alpha} \left(\sqrt{g}+2\right)\right]^{-1}}.
\end{align}
Equations \eqref{eq:al}-\eqref{eqn:pValue} with \eqref{eqn:gValue} and \eqref{eqn:cValue} represent the solution of the problem and allow us to express all quantities relevant to the problem as a function of the load $W$ (recalling that we use the contact slenderness $g$ as a parameter to facilitate plots). We now proceed to study their behaviour and compare with previous results.

\begin{figure}[t]
\begin{center}
\includegraphics[width = \textwidth]{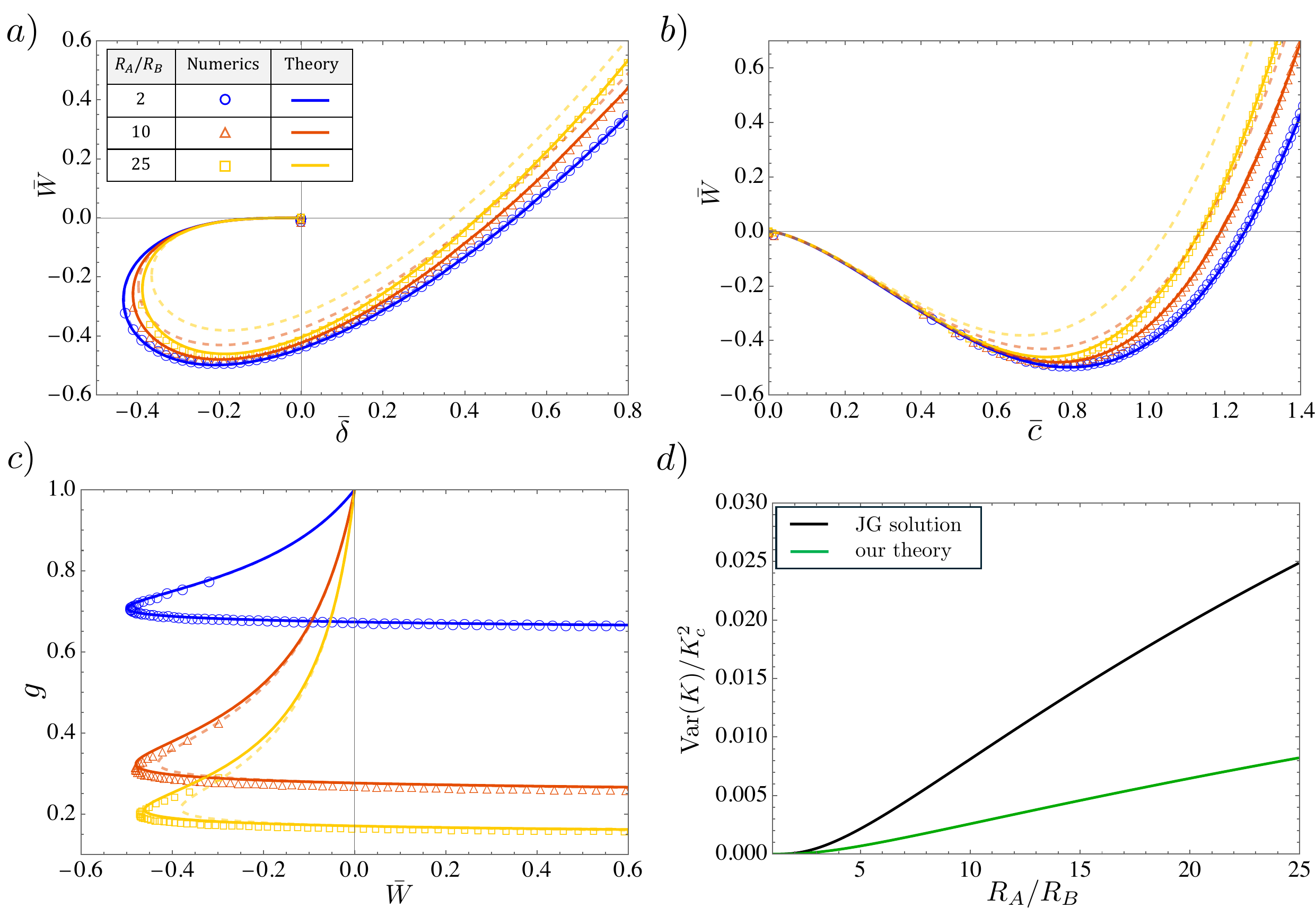}
\caption{Comparison between the present theory (solid curves), the predictions of JG (dashed curves) and recent numerical simulations (points) \cite{li2020numerical}. Plots show (a) the load versus displacement, (b) load versus contact radius, and (c) contact aspect ratio $g$ versus load for a range of values of $\lambda^2=R_B/R_A$. Since the contact region is not perfectly elliptical in numerical simulations, the aspect ratio $g$ is calculated by dividing the largest distance on the boundary from the centre (equivalent to the semi-major axis) by the smallest distance (equivalent to the semi-minor axis). d) Plot of the normalised average squared difference between the stress intensity factor $K$ and its critical value $K_c$ for different values of $R_A/R_B$ as predicted by JG's work and our theory.}
\label{fig:comparison}
\end{center}
\end{figure}

\section{Results}

We compare our theory with the results presented by JG as well as recent numerical simulations for elliptical contact obtained in \cite{li2020numerical}. To make the comparison, we express our quantities in dimensionless form following the original work by Johnson Kendall and Roberts \cite{johnson1971surface}, so that:
\begin{align}
\notag
    c_r &= \left( \frac{9 \pi R^2 \Delta \gamma}{4 \Ef} \right)^{1/3}, & \bar{W} &= \frac{W}{3\pi R \Delta \gamma}, \\
    \bar{c} &= \frac{c}{c_r}, & \delta &= \bar{\delta} \frac{c_r^2}{R}.
\end{align}
Applying these rescalings, all our equations in the previous section become independent of the parameters $E^*$, $W$, $R$ and $\delta$, and our results can be neatly presented. 
Our results are shown in figure \ref{fig:comparison}. The energy minimization approach used here leads to analytical results that are in very good agreement with numerical simulations, even for very slender contact ellipses; this good agreement is observed despite the (erroneous) assumption of an elliptical contact region. 
Note that the numerical simulations make no assumptions about the shape of the contact set, which is close to being elliptical, but slightly `squared-out' near the semi-major axis, as shown in figure 3 of \cite{li2020numerical}. 



\section{Conclusion}

In this paper, we have revisited the seminal work by Johnson and Greenwood \cite{johnson2005approximate} on non-axisymmetric adhesive contact. 
To derive our approximate theory, we maintained JG's simplifying assumption that the contact region is elliptical. However, instead of using Griffith's criterion locally to establish the effect of adhesion, we minimized the sum of the elastic and surface energies to find the size of the contact region.
The results of our theory are in good agreement with numerical simulations and offer a considerable improvement on JG's theory.

The reason for the improvement can be understood by considering the stress intensity factor at the boundary of contact. Due to the approximation of an elliptical contact, the stress intensity factor is not constant on the contact boundary. JG choose to satisfy Griffith's criterion locally, at the ends of the semi-major and semi-minor axis, where the stress intensity factor is, in fact, maximal. As a result,
the stress intensity factor $K$ differs significantly from its critical value $K_c$ away from these points, leading to a large error in the estimation of the contact size. Conversely, our global approach yields a stress intensity factor that is, on average, closer to the critical value. This improvement becomes clear when evaluating the average square difference between $K$ and $K_c$ weighted by the arclength of the ellipse, defined as
\begin{equation}
    \text{Var}(K)=\frac{\displaystyle\int_0^{2 \pi} (K-K_c)^2 \sqrt{a^2 \cos^2 \theta + b^2 \sin^2 \theta}\,\mathrm{d} \theta}{\displaystyle\int_0^{2 \pi}  \sqrt{a^2 \cos^2 \theta + b^2 \sin^2 \theta}\,\mathrm{d} \theta}.
\end{equation}
Here, $\mathrm{d}s=\sqrt{a^2 \cos^2 \theta + b^2 \sin^2 \theta}\,\mathrm{d} \theta$ is the arclength element, and we have normalised the average by the perimeter of the ellipse (so that the solution is independent of the ellipse radius, $c$). In figure \ref{fig:comparison}(d), we plot this quantity as a function of the geometric parameter $R_A/R_B$ for both JG's theory and our work. Our global energetic approach leads to a significantly smaller average error in the estimation of $K$, thus predicting the size of the contact region and the adhesive behaviour more accurately.

Using a global energetic approach instead of a local application of Griffith's criterion may help smear out errors in other problems in adhesion and contact mechanics. For example, Zini \emph{et al.}~\cite{zini2018extending} have proposed a double-Hertz model to describe elliptical contact and capture effects of adhesion also in the case of long-ranged interactions. Similarly to Johnson and Greenwood, they made a series of assumptions about the shape of the contact and cohesive region to simplify the problem---leading to a non-constant stress singularity at the boundary---and included the effects of adhesion via Griffith's criterion applied to the semi-major axis. It is likely that a part of the observed error in their theory is a result of the local application of Griffith's criterion, and we therefore anticipate that a global energetic approach may lead to more accurate results. \\\\

\section*{Acknowledgements}

A.G. is grateful to the Royal Society for funding. The project was partially supported by the UK Engineering and Physical Sciences Research Council via grant \linebreak EP/W016249/1 (D.V.).

\section*{References}
\bibliographystyle{iopart-num}
\bibliography{bib}

\end{document}